\documentclass[aps,prl,twocolumn,floatfix,superscriptaddress]{revtex4}

\usepackage{epsfig}
\usepackage{amsmath}

\begin{document}
\preprint{DUKE-TH-02-264}

\title{Photon interferometry of  
	Au+Au collisions at the Relativistic Heavy-Ion Collider}
\author{Steffen A.~Bass}
\affiliation{Department of Physics, Duke University, 
             Durham, North Carolina 27708-0305}
\affiliation{RIKEN BNL Research Center, Brookhaven National Laboratory, 
             Upton, New York 11973, USA}
\author{Berndt M\"uller}
\affiliation{Department of Physics, Duke University, 
             Durham, North Carolina 27708-0305} 
\author{Dinesh K.~Srivastava}
\affiliation{Variable Energy Cyclotron 
Centre, 1/AF Bidhan Nagar, Kolkata 700 064, India}            
\date{\today}

\begin{abstract}
We calculate the two-body correlation function of direct photons
produced in central
Au+Au collisions at the Relativistic Heavy Ion Collider.
Our calculation includes contributions from the early
pre-equilibrium phase in which photons are produced via hard 
parton scatterings as well as radiation of photons from a thermalized
Quark-Gluon-Plasma (QGP) and the subsequent expanding hadron gas.
We find that high energy photon
interferometry provides a faithful probe of the details of 
the space-time evolution and of the early reaction stages 
of the system.
\end{abstract}

\maketitle
Recent data from the Relativistic Heavy-Ion Collider (RHIC) at Brookhaven 
National Laboratory have provided strong evidence for the existence of a transient QGP 
-- among the most exciting findings are  a strong collective flow 
saturating the hydrodynamic limit \cite{v2_exp,v2_theo},
a suppression of high-$p_T$ hadrons indicative of parton energy loss
\cite{jetq_theo,jetq_exp}, and an evidence for hadronization by 
recombination of quarks from a deconfined phase \cite{reco}.
Being of hadronic nature, these probes are necessarily affected by 
the late stages of the reaction and do not provide unambiguous
information about the early equilibration phase.
Photons and dileptons, on the other hand, decouple from the system 
immediately after their creation, and thus serve as complementary 
deep probes of the hot and dense matter. Recent studies have suggested 
that rescattering of partons enhances the production of photons
\cite{bms_phot,fms}. 

One problem affecting photons is that the
inclusive photon spectrum is dominated by radiative decays of 
long-lived hadrons (especially $\pi^0$ and $\eta$). Although the 
subtraction of these sources is possible with great effort, this
problem severely limits the use of photons as penetrating
probes. Two-photon momentum correlations 
do not suffer from this problem, because the geometric size of
the distribution of 
decay vertices is extremely large on a nuclear length scale,
and thus all correlations between photons from the decay of two
different hadrons are confined to extremely small relative momenta.
This property makes photon interferometry a promising
tool for the experimental study of the pre-equilibrium dynamics
of relativistic heavy ion collisions. 

In the present work we employ the parton cascade model
(PCM) \cite{GM92,VNIBMS} and ideal hydrodynamics \cite{hydro}
to calculate two-body quantum correlations \cite{pratt} of
high energy photons and show that they can provide a sensitive test of 
the space-time evolution of the matter in the collision
zone. The peculiar features of the correlation function make it possible
to differentiate between direct photons emitted in the early pre-equilibrium
reaction phase from those produced at later times in a thermalized QGP and
subsequent hadron gas.
Following the early suggestion of Neuhauser \cite{neu}, photon 
interferometry of the QGP was investigated 
extensively by several authors \cite{dks_int,tim,pisut,slotta,bs,alam}.
However those studies focused only on photons emitted from a 
thermalized source.
Recently, interferometry measurements with photons of transverse 
momenta of 100-300 MeV/c
have been reported at SPS energies \cite{wa98}.

Here we are primarily interested in the higher-energy photons 
originating in the rapidly evolving pre-equilibrium phase when 
the initial-state partons undergo multiple inelastic collisions. 
We use the Parton Cascade Model to describe this phase of the collision.
The basic assumption underlying the PCM is that the state of 
the dense partonic system can be characterized by a set of 
one-body distribution functions $F_i(x^\mu,p^\alpha)$, where $i$
denotes the flavor index ($i = g,u,\bar{u},d,\bar{d},\ldots$)
and $x^\mu, p^\alpha$ are coordinates in the eight-dimensional
phase space. The time-evolution of the parton distribution is 
described by a relativistic Boltzmann equation, whose collision 
term includes all scattering processes involving
quarks, gluons, and photons in lowest-order quantum chromo- and 
electro-dynamics.
A low momentum-transfer cut-off $p_T^{{\rm min}}$ is introduced to 
regularize the infrared divergence of the perturbative parton-parton 
cross sections.  Additionally, we include the branchings $q \to q g$, 
$q \to q\gamma$, $ g \to gg$ and $g \to q\overline{q}$ following
two-body collisions in the leading logarithmic approximation. 
The soft and collinear singularities in the showers 
are eliminated by terminating the branchings when the virtuality 
of the time-like partons drops below $\mu_0 = 1$ GeV.  
The present work is based on the thoroughly revised, corrected, and 
tested implementation {\sc VNI/BMS} of the parton cascade 
model \cite{VNIBMS}. 
We already presented the rapidity distribution and the
transverse momentum spectra of direct photons predicted by
{\sc VNI/BMS} in an earlier publication \cite{bms_phot}. Here
we explore two-photon correlations for the same model.

\begin{figure}
\centerline{\epsfig{file=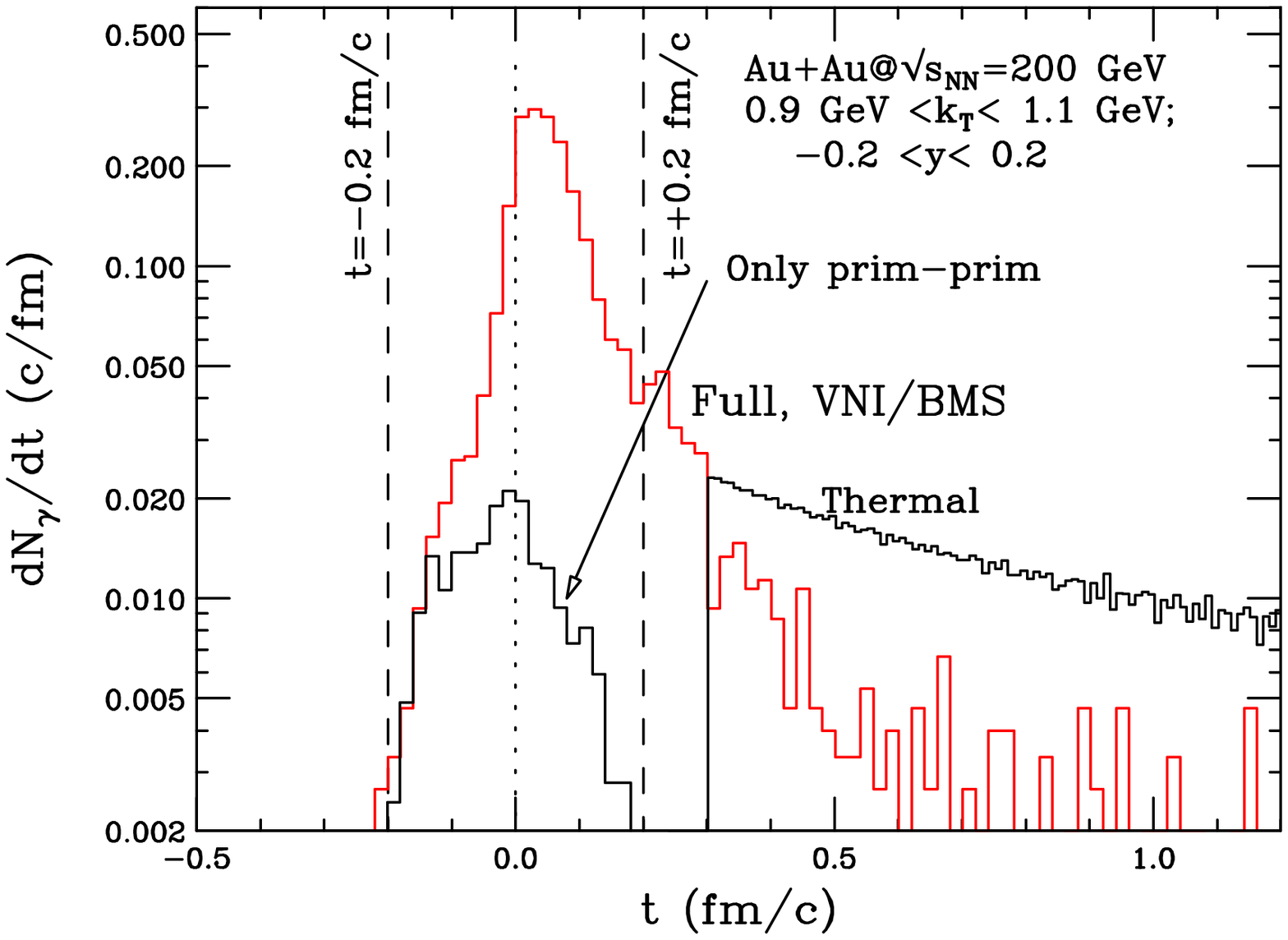,width=7.6cm}}
\centerline{\epsfig{file=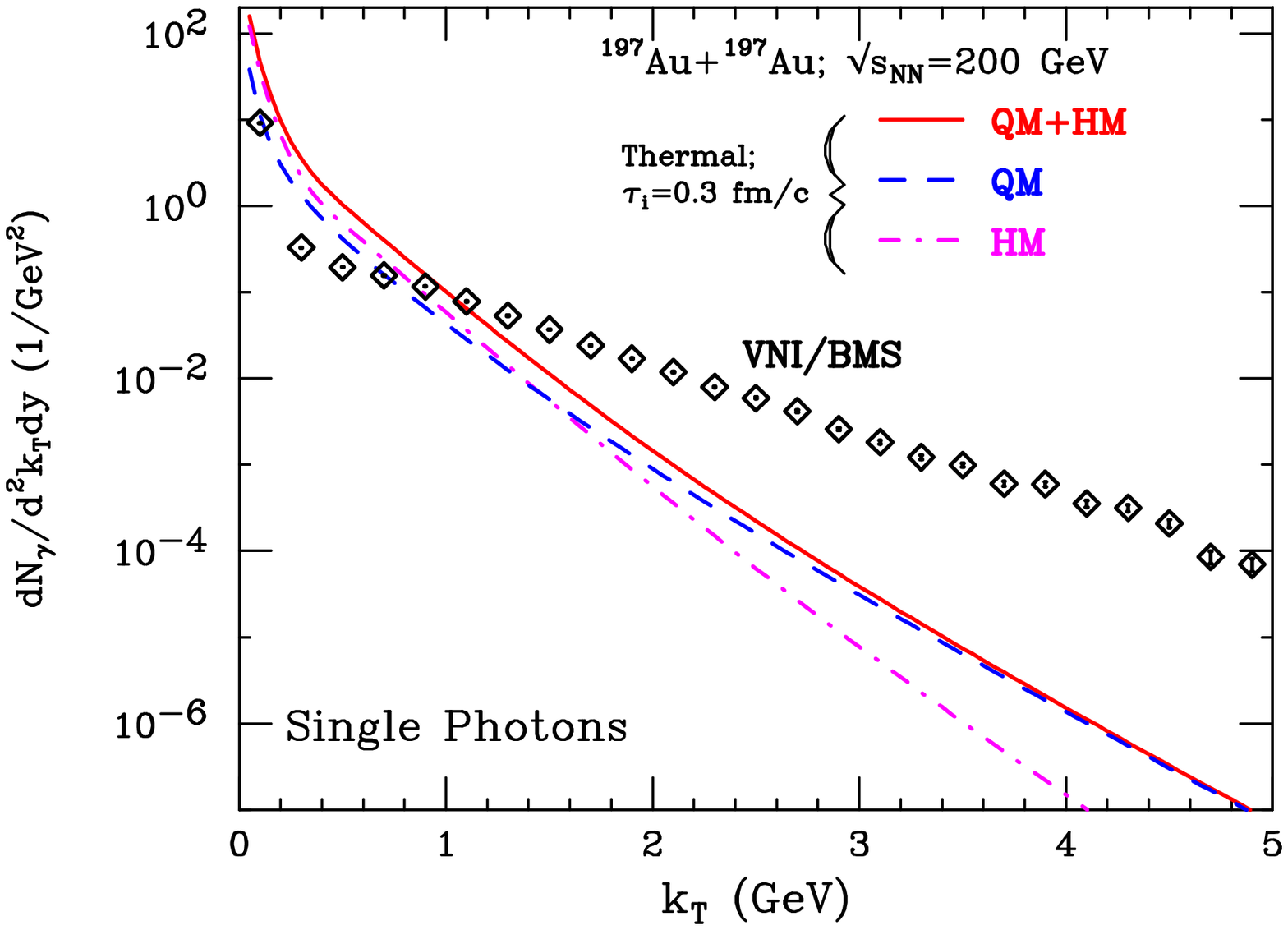,width=7.6cm}}
\caption{Upper panel: The production rate (per event) of hard photons 
         in a central collision of gold nuclei, at $\sqrt{s_{NN}}=$ 
         200 GeV as a function of time in the center-of-mass system. 
         Lower panel: Spectrum of photons from various sources:
	QM denotes photons emitted from a QGP and HM denotes photons
	emitted from a hadron gas.}
\label{fig1}
\end{figure}

The PCM calculation of the hard pre-equilibrium photons is augmented
by a hydrodynamical calculation for photon emission from a thermalized
QGP and subsequently forming hadron gas. Here, we employ a 
boost invariant  hydrodynamics \cite{hydro}, which has been used 
extensively to explore photon and dilepton production and
 also hadronic spectra. The initial conditions are estimated from 
a particle rapidity density of  about 1260 near $y=0$ for a central 
collision of gold nuclei at $\sqrt{s_{NN}}=$ 200 GeV.  We assume that
a chemically and thermally equilibrated plasma is formed at 
$\tau_0 = 0.3$ fm/$c$. We use the complete leading-order results for the 
production of photons from the QGP~\cite{guy} and the latest results 
for the radiation of photons from a hot hadronic gas \cite{simon}.

The upper frame of Figure~\ref{fig1} shows the emission rate
$dN_\gamma/dt$ for photons with a transverse momentum of $(1\pm 0.1)$ 
GeV/$c$ around mid-rapidity as a function of time.
$t=0$ is defined in the nucleus-nucleus center-of-mass frame, as 
the instant when the two Lorentz contracted nuclei completely overlap. 
88\%  of all photons produced via hard parton scattering in
the PCM are emitted within the first 0.3 fm/c of the reaction. 
Beyond this time, due to its artificial infrared cutoffs, the
PCM becomes an increasingly ineffective description of the
medium evolution and does not achieve or maintain full equilibrium
among the interacting partons. Assuming
an initial time on the order of 0.3 fm/$c$, the hydrodynamic
calculation allows for a smooth continuation of the emission rate
as a function of time for the QGP and subsequent hadron gas phase.
Note that the production of photons due to the primary-primary scatterings
only is approximately 
symmetric around $t=0$, confirming the build-up of secondary 
scatterings over time.

The relative weight of photons emitted in the pre-equilibrium phase from the
PCM vs. those stemming from a thermalized QGP and subsequent hadron gas can
be studied as a function of transverse momentum in the lower frame 
of Figure~\ref{fig1}. For a transverse momentum of $k_T = 1$~GeV/$c$ 
pre-equilibrium and thermal photons both contribute roughly 50\% to the 
total photon yield. We therefore expect both contributions to be
of importance for the calculation of the correlation function.
At $k_T = 2$~GeV/$c$, however, direct photons from the PCM outweigh thermal
photons by one order of magnitude. Here we expect only negligible 
contributions of thermal photons to the correlation function and a 
clear signal of the early photon emitting source.

The correlation between two photons with momenta $\bf{k_1}$ and $\bf{k_2}$,
averaged over their spins is

\begin{equation}
C(\mathbf{q},\mathbf{K})=1+ \frac{1}{2} \frac{\left|\int \, d^4x \, S(x,\mathbf{K})
e^{ix \cdot q}\right|^2}
                        {\int\, d^4x \,S(x,\mathbf{k_1}) \,\, \int d^4x \,
S(x,\mathbf{k_2})}
\label{eqCanal}
\end{equation}
where $S(x,\bf{k})$ is the photon source function for a 
completely chaotic source, and
\begin{equation}
\mathbf{q}=\mathbf{k_1}-\mathbf{k_2}, \,\, \,
 \mathbf{K}=(\mathbf{k_1}+\mathbf{k_2})/2\, \, .
\label{eqqK}
\end{equation}

Our assumption of a chaotic source neglects contributions to the
2 photon yield by correlated 2 photon emission. We estimate these
contributions to be small and not to influence the
shape of the correlation function.

In the parton cascade model calculation, where the probability distribution 
of the production vertices and momenta of the photons are known, 
the correlation function 
$C$ is calculated by  following the Wigner function 
scheme of \cite{uli}
using the {\sc Hansa} code \cite{sollfrank}, i.e. 
by rewriting the numerator of the second term in 
Eq.(\ref{eqCanal}) as, 
\begin{equation}
\rm{Num}(\mathbf{q}, \mathbf{K})=
 \sum_{i,j}\cos [ q \cdot (x_i-x_j) ] .
\label{eqCMC}
\end{equation}
where  the sum runs over the pairs whose individual momenta lie 
in a small bin around $\mathbf{K}$ of width
$\mathbf{\epsilon}$ in each of the four directions.

The emission vertices $(x,\mathbf{k})$ generated by a semiclassical
transport model or the monte-carlo sampling of emission rates
generally do not represent
a valid Wigner function, since the uncertainty relation
$\Delta \mathbf{x} \, \Delta \mathbf{k} > \hbar$ is not 
incorporated. In order to remove this deficiency we
have smeared the emission vertices by adding 
random fractions of $\hbar/p_i)$ to the emission vertices $x_i$
\cite{smear}.

\begin{figure}
\centerline{\epsfig{file=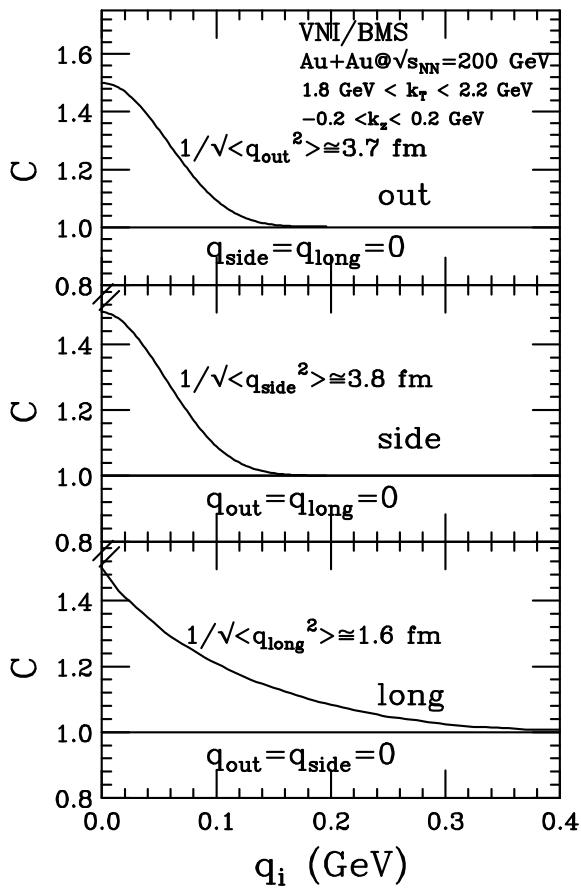,width=7.6cm}}
\caption{The outward, sideward,  and longitudinal correlations of
         direct photons predicted by the parton cascade model, at 
         $K_T=$ 2 GeV/$c$. Note the change of scales along the 
         $x$ and $y$-axes for the longitudinal correlation. 
        The inclusion of thermal
	photons changes the results only marginally.}
\label{fig2}
\end{figure}

We shall present our results for the correlation function $C(\bf{q},\bf{K})$
in terms of the so-called outward, sideward, and longitudinal momentum 
differences. We represent the photon four-momentum in the form
\begin{equation}
k^\mu=(k_{T}\, \cosh \, y,
k_{T}\,\cos \psi,k_{T} \sin \psi,k_{T} \, \sinh\, y)
\end{equation}
where $k_T$ is the transverse momentum, $y$ is the rapidity, 
and $\psi$ is the azimuthal angle.  Using the same notation for 
the relative and the average momenta (\ref{eqqK}), we can 
write~\cite{sinyukov}:
\begin{eqnarray}
q_{\text{long}} &=& |k_{1z}-k_{2z}| = |k_{1T} \sinh y_1 - k_{2T} \sinh y_2|\\
q_{\text{out}}  &=& \mathbf{q_T}\cdot \mathbf{K_T}/K_T \\
q_{\text{side}} &=& \left|\mathbf{q_T}-q_{\text{out}}
                    \mathbf{K_T}/K_T\right| \,.
\end{eqnarray}
The corresponding radii for a completely chaotic source  are 
often obtained by fitting the correlation 
function $ C(q_{\text{out}}, q_{\text{side}},q_{\text{long}})$ 
to the functional form:\\ $1+ (1/2) \cdot \exp\left[- \left( 
q_{\text{out}}^2R_{\text{out}}^2+
q_{\text{side}}^2R_{\text{side}}^2+
q_{\text{long}}^2R_{\text{long}}^2\right)/2\right] $.
Note that this provides that $\langle q_i^2\rangle =1/R_i^2$, 
where $i$=out, side, long.

Transport calculations have employed several procedures to improve the
significance of their results in order to compare with experimental data. 
 Either the source function is parametrized, permitting
an analytical evaluation of eq.\ (\ref{eqCanal}), or the particles 
generated in the calculation are ``dispersed'' over many particles 
to enhance the pair statistics. For our analysis we 
have accumulated 125000 events from VNI/BMS
to make an accurate determination of the source radii possible 
by direct evaluation of eq.\ (\ref{eqCMC}). We supplement
these results with photons from a Monte-Carlo sampling of
the hydrodynamics calculation to obtain the production vertices and
momenta of the photons, which are combined with those from the PCM 
in our correlation analysis.
 
In Fig.~\ref{fig2} we show our results for photon momenta of
 $k_T = (2 \pm 0.2)$~GeV, 
where the high energy photons stemming from hard parton scattering 
completely dominate the total photon yield. 
We study the three-dimensional correlation function, in terms of one 
momentum coordinate, limiting the other two
momentum coordinates to zero. The sideward 
radius as inferred from photons is $R_{\rm side}\approx 3.8$~ fm,
while the corresponding outward radius is found to be 
$R_{\rm out}\approx 3.7$~fm. These values are somewhat smaller than the Gaussian 
equivalent radius of the density distribution of a Au nucleus, reflecting
the geometric shape of the colliding nuclei. The very small difference 
between the two radius parameters indicates a short source lifetime. 

The longitudinal correlation radius is much smaller, 
$R_{\rm long}\approx 1.6$~fm.
This is not unexpected 
\cite{dks_int} and has its origin in the fact that these 
photons are mostly emitted when the longitudinal extension of the system 
is quite small and the velocity gradients are large. 
 We have verified that $C(q_\text{long})$ is robust with respect
to reasonable variations of the center of the rapidity window
and its width, which determines $\epsilon$.
A measurement of this value 
of $R_{\rm long}$  
would clearly identify the
source of these photons as prehadronic.

\begin{figure}
\centerline{\epsfig{file=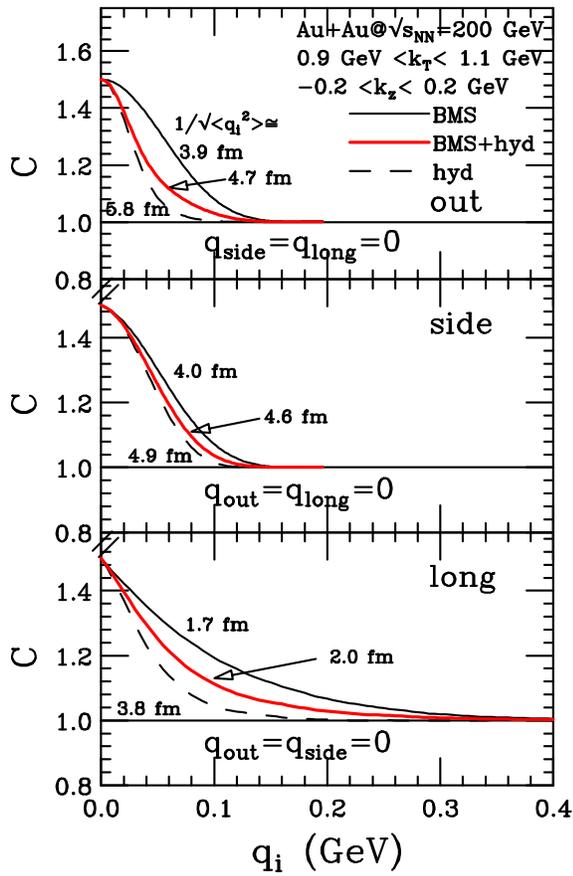,width=7.5cm}}
\caption{Outward, sideward and longward intensity correlation of photons at 1 GeV,
         considering only PCM (BMS), only thermal (hyd), and  PCM+thermal 
         photons (BMS+hyd).}
\label{fig4}
\end{figure}

In figure~\ref{fig4} we show our results for photons at
$k_T = (1 \pm 0.1)$~GeV. For this choice of $k_T$ pre-equilibrium 
photons and thermal ones from the QGP and hadron gas phases of the 
reaction contribute about equally to the yield -- therefore both 
contributions need to be taken into account when calculating the 
correlation functions. As is to be expected, the radii
extracted from the correlation functions lie between those
of the individual
contributions (thermal photons vs. pre-equilibrium photons), 
however, they in general do not reflect the average of 
the radii
as the relative single photon yields would suggest, due to
the influence of the cross-term, i.e. the interference of 
thermal photons with pre-equilibrium photons.

For low $k_T$, thermal photons should completely dominate:
at $k_T$ = 0.2 GeV, 
our calculations predict $R_{\rm side}$ of
about 5 fm, with $R_{\rm out}/R_{\rm side}\approx$ 2.

In summary, we have calculated the correlation functions of high energy
photons  emanating from the pre-equilibrium stage as well as the thermal
QGP and hadron gas stages  of the matter formed in central 
Au+Au collisions 
at RHIC. Our study predicts 
that photon interferometry 
will reveal 
a small source
of brief duration for photons at transverse 
momenta $k_T \ge 2$~GeV/$c$ due to pre-equilibrium emission. 
This contrasts with a much larger source, 
revealed by thermal photons at lower momenta, as they mostly originate 
from the QGP and hadron gas stage when the system has already expanded
significantly.

\begin{acknowledgments}  
This work was supported in part by RIKEN, the Brookhaven National 
Laboratory, DOE grants DE-FG02-96ER40945 and DE-AC02-98CH10886. 
We thank J. Sollfrank for providing us with the {\sc Hansa} code.
\end{acknowledgments}

\end{document}